\documentclass[prd,twocolumn,superscriptaddress,floatfix,amsmath,amssymb,amsfonts,longbibliography,nofootinbib]{revtex4-2}

\usepackage{float} 

\usepackage[normalem]{ulem}
\usepackage[english]{babel}
\usepackage{graphicx}
\usepackage{dcolumn}
\usepackage{bm}
\usepackage{blindtext}
\usepackage{verbatim}
\usepackage{relsize}
\usepackage{mathrsfs}
\usepackage{musicography}
\usepackage{amsmath}
\usepackage{blindtext}
\usepackage{cancel}
\usepackage{physics}
\usepackage{epstopdf}
\usepackage{mathtools}
\usepackage{blindtext}
\usepackage{tensor}
\usepackage{color}
\usepackage[usenames,dvipsnames]{pstricks}
\usepackage{epsfig}
\usepackage{pst-grad} 
\usepackage{pst-plot} 
\usepackage{hyperref}
\usepackage{verbatim}
\usepackage{slashed}
\usepackage{dsfont}

\allowdisplaybreaks[1] 

\begin{document}

\title{Carrollian Motion in Magnetized Black Hole Horizons}


\author{Finnian Gray}
\email{fgray@perimeterinstitute.ca}

\affiliation{Department of Physics and Astronomy, University of Waterloo, Waterloo, Ontario, N2L 3G1, Canada}
\affiliation{Perimeter Institute for Theoretical Physics, Waterloo, Ontario, N2L 2Y5, Canada}

\author{David Kubiz\v{n}{\'a}k}
\email{david.kubiznak@matfyz.cuni.cz}

\affiliation{Institute of Theoretical Physics, Faculty of Mathematics and Physics,
Charles University, Prague, V Hole\v{s}ovi\v{c}k\'ach 2, 180 00 Prague 8, Czech Republic}

\author{T. Rick Perche}
\email{trickperche@perimeterinstitute.ca}

\affiliation{Department of Applied Mathematics, University of Waterloo, Waterloo, Ontario, N2L 3G1, Canada}
\affiliation{Perimeter Institute for Theoretical Physics, Waterloo, Ontario, N2L 2Y5, Canada}
\affiliation{Institute for Quantum Computing, University of Waterloo, Waterloo, Ontario, N2L 3G1, Canada}

\author{Jaime Redondo--Yuste}
\email{jaime.redondo.yuste@nbi.ku.dk}

\affiliation{Niels Bohr International Academy, Niels Bohr Institute, Blegdamsvej 17, 2100 Copenhagen, Denmark}

\begin{abstract}
We revisit the motion of massless particles with anyonic spin in the horizon of Kerr--Newman geometry. As recently shown, such particles can move within the horizon of the black hole due to the coupling of charges associated with a 2-parametric central extension of the 2-dimensional Carroll group to the 
magnetic field generated by the black hole -- the so called ``anyonic spin-Hall effect''. 
We show 
that the previously computed
magnetic field is not invariant under Carroll diffeomorphisms and find another result which respects these symmetries of the horizon. We also consider a more astrophysically relevant case of a (weakly charged) rotating back hole placed in a uniform magnetic field, which could, for instance, be induced by the surrounding plasma. We show that a qualitatively similar magnetic field assisted anyonic spin--Hall effect takes place, even in the absence of black hole rotation. The theoretical possibility of a motion induced by a magnetic monopole is also studied.
\end{abstract}

\maketitle

\section{Introduction}

Black holes are among the most fascinating objects in our Universe. 
Their fundamental characteristic is the existence of the {\em black hole horizon} where many interesting physical phenomena 
are predicted to happen. Among these, the most puzzling is the quantum black hole evaporation~\cite{Hawking:1975vcx}, and the associated black hole information paradox \cite{Hawking:1976ra, Page:1993wv, Page:2013dx, Almheiri:2020cfm}. 
From a classical perspective, the event horizon is the surface which defines the point of no return for infalling observers. In fact, any massive particle would require infinite proper acceleration to remain on the horizon. However, the event horizon is also a {\em null surface}, which implies that, in principle, a massless particle could stay on the horizon without falling into the black hole.

The symmetry group of a null surface is the {\em Carroll group}, which can be thought of as the vanishing speed of light limit, $c\rightarrow 0$, of the Poincar\'e group~\cite{Dautcourt:1997hb,Duval_2014,Ciambelli:2019kiw}.  The dynamics of particles in Carroll manifolds can be formulated in an associated Carroll-homogeneous symplectic manifold~\cite{first}. These so-called {\em Carroll particles} then have unique dynamics that is a consequence of the particular group structure that defines them. This procedure can also be carried out in the horizon of a black hole, and the associated dynamics of Carroll particles living in the horizon can be studied.

Although it has been well known that Carroll fluids have non-trivial dynamics~\cite{Bergshoeff:2014jla,Freidel:2022bai,Freidel:2022vjq}, until recently, it was thought that a single Carroll particle could not move~\cite{levy1965nouvelle,VD1966,Bergshoeff:2014jla,Duval_2014}. This belief was disproven in the case of (2+1) Carrollian manifolds in~\cite{first}, where it was shown that by considering the {\em double central extension} of the Carroll group, one endows a Carroll particle with two central charges. This allows a massless chargeless particle with anyonic spin to move under the influence of a {\em magnetic field} in a Carroll manifold. 

The results of~\cite{first} were first applied to black holes in~\cite{main,main2}, where the authors found an analogue to the spin-Hall effect for massless Carrollian particles in the horizon of a Kerr--Newman black hole, which they termed the {\em anyonic spin-Hall effect}. The induced magnetic field of the charged rotating black hole couples to the Carroll particle, and results in non-trivial motion in the horizon. However, as we will show in this manuscript, the method used in~\cite{main,main2} to compute the magnetic field induced in the horizon gives a field that does not respect the Carroll symmetry, making it incompatible with the application of the results of~\cite{first}.

In this manuscript we find a consistent way of inducing an electromagnetic tensor in the horizon, which respects its induced Carrollian symmetries. This allows us to restate the results of~\cite{main,main2}, with another expression for the resulting magnetic field. We also apply the double central extension of the Carroll group to the dynamics of anyonic particles in the horizon of astrophysically more relavant spacetimes of weakly charged rotating black holes placed in an external magnetic field \cite{wald:1974np}, obtaining so the `assisted' anyonic spin-Hall effect.

Our manuscript is organized as follows. In Section \ref{sec:KNCarroll} we review the basic properties of the horizon of a Kerr--Newman black hole, and discuss its Carrollian structure. In Section~\ref{sec:CarrollParticles} we apply the results of~\cite{first} to the motion of a massless anyonic particle in this  horizon, 
offering an alternative to the magnetic field employed in~\cite{main,main2}. In Section \ref{sec: Ind Car Mot} we show that the new magnetic field  respects the Carroll symmetries of the horizon,
favouring it against the magnetic field obtained in~\cite{main,main2}. 
In Section~\ref{sec:assisted} we consider the motion of a Carrollian particle in a rotating black hole in an external magnetic field. Section~\ref{app:magneticCharge} deals with a possibility of including black holes with a magnetic monopole. 
The conclusions of our work can be found in Section~\ref{sec:conclusions}.

\section{The Horizon of Kerr--Newman Black Holes}\label{sec:KNCarroll}

The Kerr--Newmann solution \cite{newman1965metric} describes charged and rotating black holes in Einstein--Maxwell theory.  The corresponding metric in Kerr-like coordinates $(v,r,\theta,\phi)$ is given by
\begin{align}\label{eq:KNmetric}
    \dd s^2 = &- \frac{\Delta}{\Sigma}\left(\dd v - \frac{\Sigma}{\Delta}\dd r - a \sin^2\theta \, \dd \phi\right)^2+ \frac{\Sigma}{\Delta} \dd r^2 \\&+ \frac{\sin^2 \theta}{\Sigma}\left(a \dd v - (r^2 + a^2) \dd \phi\right)^2 + \Sigma\,\dd \theta^2 ,\nonumber
\end{align}
where   
\begin{align}
    \Sigma &= r^2 + a^2 \cos^2\theta\,,\nonumber\\
    \Delta &= r^2 + a^2 + Q^2 - 2M r\,.
\end{align}
$M$ describes the black hole mass, $Q$ is its charge, and $J=Ma$ is the hole's angular momentum, written in terms of the rotation parameter $a$.
The metric admits two Killing vectors: $\partial_v$ and $\partial_\phi$, corresponding to time translations and angular rotations along the axis of angular momentum of the black hole. 

The metric \eqref{eq:KNmetric} solves the Einstein equations $G_{\mu\nu} = 8\pi T_{\mu\nu}$, where
\begin{equation}
    T_{\mu\nu} = -\frac{1}{4\pi}\Bigl(F_{\mu\alpha}F^{\alpha}{}_\nu+\frac{1}{4}g_{\mu\nu}F_{\alpha\beta}F^{\alpha\beta} \Bigr)
\end{equation}
is the stress-energy momentum tensor of the electromagnetic field
$F_{\mu\nu} = \partial_\mu A_\nu - \partial_\nu A_\mu$\,, derived from the following gauge potential:
\begin{equation}\label{eq:A}
    A = - \frac{Qr}{\Sigma}\left(\dd v - a\sin^2\theta\,\dd \phi\right)\,.
\end{equation}

The field strength $F_{\mu\nu}$ obeys the vacuum Maxwell equations $\nabla_\mu F^{\mu\nu}=0$. More explicitly, it can be written as follows:
\begin{align}
    F = &\frac{Q}{\Sigma^2}(r^2 - a^2\cos^2\theta)\dd r \wedge (\dd v - a \sin^2 \theta\,\dd \phi) \nonumber\\
    & - \frac{2 a Q r}{\Sigma^2}\sin\theta \cos \theta\, \dd\theta\wedge \Bigl(a\dd v - (r^2 + a^2)\dd \phi\Bigr)\,.
\end{align}

The black hole {\em horizon} is located at the largest root of $\Delta(r)=0$, at 
$r_+=M + \sqrt{M^2 - a^2 - Q^2}$. When $M^2=a^2+Q^2$, the horizon becomes degenerate, and the black hole is called extremal.
The chosen coordinates are regular on the horizon (the apparently singular $\dd r^2$ terms cancel out). 
The horizon is the (2+1) null surface generated by the spacelike vectors $\partial_\theta$, $\partial_\phi$ and the null vector \mbox{$\xi = \partial_v + \Omega_+ \partial_\phi$}, where
\begin{equation}
\Omega_+=\frac{a}{r_+^2+a^2}\,,
\end{equation}
is the angular velocity of the horizon. 

Taking into account the angular velocity of the Horizon, one can perform a change of coordinates, which takes this rotation under consideration. We define
\begin{equation}
    \varphi  =  \phi - \Omega_+ v 
\end{equation}
so that the metric in the new coordinates $(v,r,\theta,\varphi)$ reads
\begin{align}\label{eq:KNmetricNew}
    \dd s^2 = &- \frac{\Delta}{\Sigma}\left(\frac{r_+^2 +a^2 \cos^2 \theta}{r_+^2 + a^2}\dd v + \frac{\Sigma}{\Delta}\dd r - a \sin^2\theta \, \dd \varphi\right)^2\nonumber\\& + \frac{\sin^2 \theta}{\Sigma}\left(a\left(\frac{r_+^2-r^2}{r_+^2 + a^2}\right) \dd v - (r^2 + a^2) \dd \varphi\right)^2 \nonumber\\&+ \frac{\Sigma}{\Delta} \dd r^2+ \Sigma\,\dd \theta^2.
\end{align}
In these coordinates, the vector $\partial_v$ is then tangent to the horizon and null at $r=r_+$. The electromagnetic potential reads
\begin{equation}
    A =  - \frac{Qr}{\Sigma}\left(\frac{r_+^2 +a^2\cos^2\theta}{r_+^2 + a^2}\dd v - a \sin^2 \theta \,\dd \varphi\right),
\end{equation}
and the electromagnetic field can be written as
\begin{align}
    F = &\frac{Q}{\Sigma^2}(r^2 - a^2\cos^2\theta)\dd r \wedge \left(\frac{r_+^2 +a^2\cos^2\theta}{r_+^2 + a^2}\dd v - a \sin^2 \theta\,\dd \varphi\right) \nonumber\\
    & - \frac{2 a Q r}{\Sigma^2}\sin\theta \cos \theta\, \dd\theta\wedge \left(a\left(\frac{r_+^2 -r^2}{r_+^2 + a^2}\right)\dd v - (r^2 + a^2)\dd \varphi\right)\,.
\end{align}

One can then obtain a null form normal to the horizon, $n_\mu$ which is orthogonal to $\partial_\theta$ and $\partial_r$ and such that $n_\mu \xi^\mu = 1$. It is explicitly given by
\begin{equation}
    n = \dd v -\frac{a(r_+^2 + a^2)}{\Sigma_+} \sin^2 \theta \,\dd\varphi,
\end{equation}
where $\Sigma_+$ denotes $\Sigma$ evaluated at $r = r_+$. Using $\xi$ and $n$ one can define the projector into the horizon as
\begin{equation}\label{projectorq}
 q^\mu{}_{\nu} = \delta^\mu_\nu - \xi^\mu n_\nu - n^\mu \xi_\nu\,,    
\end{equation}

or, in terms of the basis of 1-forms,
\begin{equation}\label{eq:induced-metric}
    q = \frac{(r_+^2 + a^2 )^2\sin^2 \theta}{\Sigma
    _+}  \dd \varphi^2 + \Sigma_+ \dd\theta^2.
\end{equation}
The latter represents a degenerate metric on a $(2+1)$ surface. The horizon $\mathcal{H}$ is therefore a null $2+1$ surface, but with a degenerate metric. Therefore we cannot study it as any other (pseudo)-Riemannian manifold embedded in spacetime. Recently it was shown~\cite{Donnay:2019jiz} that black hole horizons can be endowed with a Carrollian structure~\cite{Bergshoeff:2022eog}. This structure not only allows for a consistent treatment of null surfaces, but also makes explicit the symmetries of black hole horizons. 

The starting point is constructing coordinates $(v,\rho,\theta,\varphi)$, valid sufficiently close to the horizon, such that the metric acquires a simple form~\cite{chrusciel2020geometry}. The construction procedure is carefully described for general spacetimes in~\cite{Booth:2012xm} and then specified to the case of the Kerr--Newman black hole. The metric in these new coordinates is:
\begin{equation}\label{eq:metric}
    \begin{aligned}
        ds^2 = &-\rho \alpha^2 \dd v^2 + 2\dd v \dd \rho - 2\rho U_A \dd v \dd x^A + \\
        &+(\Omega_{AB} - \rho \lambda_{AB})\dd x^A \dd x^B + \mathcal{O}(\rho^2),
    \end{aligned}
\end{equation}
%
where, denoting $\chi_+\equiv r^2_+ +a^2$ and $\Delta'_+=\Delta'(r_+)=2\sqrt{M^2-a^2-Q^2}$, we have
\begin{equation}
    \begin{aligned}
        \alpha^2 &= \frac{\Delta'_+}{\chi_+}\,, \quad U_\theta = \frac{2a^2\sin\theta\cos\theta}{\Sigma_+}, \\
        U_\varphi &= \frac{a\sin^2\theta}{\Sigma_+^2}\left[2r_+\chi_++ \Sigma_+\Delta'_+\right], \\
        \Omega_{\theta\theta} &=\Sigma_+\,,\quad\Omega_{\varphi\varphi}=\frac{\chi_+^2}{\Sigma_+}\sin^2\theta\,, 
    \end{aligned}
\end{equation}
and finally $\lambda_{AB}$ is given by
\begin{align}
    \quad \lambda_{\theta\theta} &=\frac{2 r_+\chi_+}{\Sigma_+}\,,\quad \lambda_{\theta\varphi}=-\frac{2\chi_+a^3\sin^3\!\theta\cos\theta}{\Sigma_+^2}\,,\nonumber\\
    \lambda_{\varphi\varphi}&=\chi_+\sin^2\!\theta\frac{2r_+\chi_+\Sigma_+ -a^2\sin^2\!\theta(2r_+\chi_+ +\Delta'_+\Sigma_+) }{\Sigma_+^3}\,.
\end{align}
In these coordinates the horizon is located at $\rho=0$, where $\rho$ is the affine distance to the horizon. Notice that the induced metric at the horizon $q$ is obtained in this coordinates by just setting $\rho = 0$, where it is explicitly degenerate. The coordinate {$v$} is the advanced Eddington-type coordinate, which acts as a ``null clock'' in the horizon. The coordinates $x^A=\{\theta,\varphi\}$ parameterize the angular direction in the $2$-dimensional slices of the horizon at fixed {$v$}.

The gauge potential can also be written as a Taylor expansion in $\rho$ in these coordinates. To leading order in $\rho$ its components are explicitly given by~\cite{Booth:2012xm,Setare:2018ziu}:
\begin{align}\label{eq:Aexp}
        &A_v = - \frac{Qr_+}{r_+^2 + a^2}+O(\rho), \quad A_\rho = O(\rho^2), \quad A_\theta = O(\rho), \nonumber\\
        &A_\varphi = \frac{aQ\,r_+\sin^2\theta}{r_+^2+a^2\cos^2\theta}+O(\rho).
\end{align}

There is a natural class of diffeomorphisms that act on the 2+1 dimensional horizon, defining its symmetry group: the Carroll group~\cite{Donnay:2016ejv}. These can be obtained as the projection of spacetime diffeomorphisms into the horizon. The infinitesimal generators of the Carroll algebra can be written as
\begin{equation}
    \chi = f({v}, x^A)\partial_{{v}} + Y^A(x^B)\partial_A,
\end{equation}
for an arbitrary super-translation $f$ and a super-rotation $Y^A$. The Carroll group appears naturally~\cite{levy1965nouvelle, Duval_2014} when one considers the contraction of the usual Poincar\'e group as the speed of light goes to zero, $c \to 0$. In that sense, the Carroll group is dual to the Galileo group, where the concepts of space and time get interchanged. The remarkable observation is that this group also appears as the symmetry group of the metric~\eqref{eq:metric} as one takes surfaces with constant $\rho$, in the limit where $\rho \to 0$, \emph{i.e.} in the horizon of a black hole. We will use the fact that the coordinate $\rho$ plays a similar role in the gravitational sector to the one played by the speed of light in Minkowski space when we discuss the induced magnetic field in the horizon of a black hole.

Now we show explicitly that the horizon $\mathcal{H}$ can be endowed with a Carroll structure. This structure is given by a fibre bundle $p$: $\mathcal{H}\to S$, where $S$ has the topology of $S^2$ and is just any constant $v$ slice,  $S = \mathcal{H}\rvert_{v = v_0}$. The projection is the usual projector constructed from the metric, and the surface $S$ is a Riemannian manifold with a metric given by 
\begin{equation}\label{2dmetric} 
\dd l^2 = \Omega_{AB}\dd x^A \dd x^B
=\sum_{{\hat A}=1,2}(e^{\hat A})^2\,,
\end{equation}
where the latter represents the orthonormal frame.
The fibre bundle has a vertical vector field given by $\partial_{{v}}$, which generates the vertical space (the ``time'' evolution). In order to have a full Carroll structure a preferred horizontal subspace is needed in order to define a connection which is covariant under Carrollian diffeomorphisms. In this case, the covariant derivative is given (in a schematic way) by
\begin{equation}
    \nabla_A \mapsto \hat{\nabla}_A = \hat{\partial}_A + \hat{\Gamma}_A,
\end{equation}
where $\hat{\Gamma}_A$ are the Carroll-Levi-Civita symbols and the partial derivatives get corrected as~\cite{Ciambelli:2019kiw,Booth:2012xm}
\begin{equation}
    \hat{\partial}_A = \partial_A + \frac{U_A}{\alpha^2}\partial_{v}.
\end{equation}
In the following, both the background metric and the fields are static (in $v$), so the Carroll-covariant derivative coincides with the usual covariant derivative. We refer the interested reader to~\cite{Ciambelli:2019kiw} for a complete exposition on the emergence of the Carroll structure on null surfaces.

\section{Motion of Carrollian Particle in the Horizon}\label{sec:CarrollParticles}

In~\cite{first} it was shown that a massless spinless particle confined in a (2+1) Carrollian manifold admits a non-trivial equation of motion in the presence of magnetic field. In a $(2+1)$-dimensional manifold equipped with coordinates $x^a=(v, x^A)$, the electromagnetic tensor can be decomposed  
as
\begin{equation}\label{decompose}
\tilde F=E_Adv\wedge dx^A+\hat F\,,    
\end{equation}
where 
\begin{equation}\label{B}
\hat F=\frac{1}{2}F_{\hat A\hat B}e^{\hat A}\wedge e^{\hat B}=Be^{\hat 1}\wedge e^{\hat 2}\,. 
\end{equation}
Whereas $E_A$ can change under the Carrollian diffeomorphisms, 
\begin{equation}\label{B2}
B=\frac{1}{2}\epsilon^{AB}\hat F_{AB}
\end{equation}
is a {\em scalar} under the action of the Carroll group (including the possible rotation of the orthonormal frame).

In~\cite{first} the authors show that the equations of motion for the massless spinless particle undergoing a trajectory $x^A(v)$\,, 
$A = 1,2$ are
\begin{equation}\label{eq:EoM}
    \dv{x^A}{v} =  \frac{\mu \chi}{\kappa_{\text{mag}}} \epsilon^{AB}\partial_B B,
\end{equation}
where $v$ is the Carrollian time, $\mu$ is the magnetic moment of the particle, $\chi$ its anyonic spin and $\kappa_{\text{mag}}$ is a central extension parameter which allows for the particle to couple to  electromagnetism. $\epsilon^{AB}$ is the Levi-Civita tensor of the 2-dimensional metric~\eqref{2dmetric} (including its determinant: $\epsilon^{12}=1/\sqrt{\det \Omega}$). 
This equation of motion can be compared to the (3-dimensional) equation of motion describing the Spin-Hall effect~\cite{hirsch1999spin}:
\begin{equation} \label{eq:analSpinHall}
    \dv{x^i}{v} = -e (E \times \Theta)^i\,,
\end{equation}
%
where $e$ is the electric charge of the spinning particle, $\Theta$ its Berry phase, and $E$ the external electric field. If Eq.~\eqref{eq:analSpinHall} is restricted to a plane with $\Theta$ normal to it, we recover an analogue of Eq.~\eqref{eq:EoM} for the motion of an exotic photon by identifying the electric charge $e$ with the combination $\mu\chi$, the electric field $E$ with the gradient of the magnetic field $\partial_i B$, and the magnetic field responsible for the Berry phase with $\kappa_{\text{mag}}$~\cite{main,main2}.

The results above show that it is, at least in principle, possible for a massless anyonic particle to live in the horizon, and undergo a non-trivial motion due to the magnetic field of the black hole. In fact, this is what was explored in~\cite{main,main2}. In order to obtain the equations of motion for such particle, it is enough to find the induced \emph{scalar} magnetic field in the horizon. In~\cite{main,main2} the authors find
\begin{equation}\label{eq:Their-B}
    B_{\text{ref}} = \frac{2aQr_+(r_+^2+a^2)\cos\theta}{(r_+^2 +a^2 \cos^2 \theta){}^3} = (\star F)^r{}_v \, ,
\end{equation}
and thus obtain the corresponding equation of motion from Eq.~\eqref{eq:EoM}. This magnetic field can be obtained via the operation
\begin{equation}
    B_{\text{ref}} = \frac{1}{\sqrt{-g}}F_{\theta\varphi}\Big|_{r=r_+}=\frac{F_{\theta\varphi}}{\Sigma \sin\theta }\Big|_{r=r_+}\,.
\end{equation}
However, one could argue that perhaps a more natural way of obtaining such a magnetic field is by projecting the electromagnetic tensor in the horizon, using the projector $q^\mu{}_{\nu}$,~\eqref{projectorq}. We define the projected electromagnetic tensor $\tilde{F}$ as
\begin{equation}
   \tilde{F}_{ab} = q^\mu{}_aq^\nu{}_b F_{\mu\nu}\,.
\end{equation}
Notice that we have written $\tilde{F}$ as the projection of the electromagnetic tensor in the (2+1)-dimensional horizon, taking into account both its spacelike and lightlike directions. We are interested in this projection because it is the (2+1)-dimensional horizon which can be regarded as a Carrollian manifold. We obtain
\begin{equation}\label{projectedF}
    \tilde{F} = \frac{2 Q a r_+(r_+^2 + a^2) \sin\theta\cos\theta }{(r_+^2 +a^2 \cos^2 \theta){}^2}\dd \theta \wedge\dd \varphi.
\end{equation}
To extract $B$ we compare this to~\eqref{decompose}. Namely, we write the metric~\eqref{2dmetric} in an orthonormal frame
\begin{equation}
e^{\hat 1}=e^{\hat \theta}=\sqrt{\Sigma_+}\dd\theta\,,\quad e^{\hat 2}=e^{\hat \varphi}= \frac{(r_+^2+a^2)\sin\theta}{\sqrt{\Sigma_+}}\dd\varphi\,.   
\end{equation}
Using \eqref{B2} then yields 
\begin{equation}\label{eq:Our-B}
    B = 
    \frac{2a Q r_+ \cos\theta}{(r_+^2 +a^2 \cos^2 \theta)^2}\,.
\end{equation}
This is nothing else than $F_{\theta\varphi}$ evaluated on the horizon and written in the orthonormal frame.


The two magnetic fields obtained in Eqs.~\eqref{eq:Their-B} and~\eqref{eq:Our-B} are obviously different. 
(While in both cases the direction of the magnetic field is the same, in~\cite{main,main2} the magnitude of the magnetic field is larger than in our case for all values of $\theta$.)
However, there should be no ambiguity in prescribing a physical quantity that directly impacts the dynamics of a particle. In order to address which prescription of the magnetic field is correct, one requires to study in detail the compatibility of electromagnetism in $(2+1)$ with the symmetries of a Carroll manifold used to derive the equations of motion in~\cite{first}.

\section{Induced Carrollian electromagnetism}\label{sec: Ind Car Mot}

Previously we have discussed that the horizon can be endowed with a Carroll structure. The structure emerges as the $\rho\to 0$ limit in an adapted set of coordinates. This limit turns out to be equivalent to the $c\to 0$ limit of the Poincare group. This identification breaks the Lorentz invariance: the quantities that are well-defined in the Carroll structure must transform covariantly under Carrollian diffeomorphisms, and not under the whole group of diffeomorphisms. Therefore, the symmetries at the horizon give a natural prescription to define the induced magnetic field, which is the object of our interest. We will first discuss how to obtain this for a usual electromagnetic theory defined on $(2+1)$-dimensional Minkowski space, where the Carroll limit appears as the limit in which the speed of light vanishes. Then we will make use of the analogy that connects this limit to the near-horizon limit in order to extend the construction to obtain a consistent electromagnetic theory at the horizon. 

Let $A_\mu$ be the electromagnetic potential in Minkowski spacetime, which we can decomposed as $(A_t/c, A_A)$. These quantities can be expanded in powers of the speed of light as 
\begin{equation}\label{eq:EM-expansion}
    A_t = c^p \sum_{n=0}^\infty A_t^{(n)}c^{2n}, \quad A_{A} = c^p \sum_{n=0}^\infty A_{A}^{(n)}c^{2n}.
\end{equation}
for some positive constant $p$. In~\cite{deBoer22} it was proved that only the leading order term of the expansion, $A^{(0)} = (A_t^{(0)}/c, A_A^{(0)})$, is invariant under Carroll diffeomorphisms. Indeed, it is not hard to check that Carrollian boosts will mix different powers of the speed of light. Since we know that the induced electromagnetic field in the Carrollian structure (which arises as the $c\to 0$ limit) must be Carroll-covariant, it must be given by the $A^{(0)}$ sector. Notice that there is another possibility to take the $c\to 0$ limit and obtain a consistent Carrollian electromagnetism, dubbed \emph{electric} contraction in~\cite{Henneaux:2021yzg} (as opposed to the contraction taken here, referred to as \emph{magnetic} contraction). That contraction, however, involves re-scaling the fields with powers of $c$ (see below).  


Now we can discuss the electromagnetic field around a charged rotating black hole. Using the analogy between the roles played by $\rho$ in this case and $c$ in the discussion above, we consider the expansion of Eq. \eqref{eq:Aexp}. 
%
%
We notice that this expansion is precisely the expansion~\eqref{eq:EM-expansion} considering $\rho$ instead of $c$. Therefore, the Carroll-covariant sector is just given by the following contribution:
\begin{equation}\label{eq:A0}
    A^{(0)} =  \frac{aQ\,r_+\sin^2\theta}{r_+^2+a^2\cos^2\theta}\dd\varphi.
\end{equation}
The $\rho$ component must vanish since we are taking the projection onto the horizon\footnote{\label{ftnt: foliation}In this sense the structure does not depend on any foliation of stretched horizons. This is just a convenient tool that allows us to choose a particular set of coordinates where the Carroll structure becomes more intuitive.}. {Indeed, setting $\rho = 0$ in Eq.~\eqref{eq:A}, we obtain Eq. \eqref{eq:A0} for the gauge potential apart from a pure gauge term in the $\dd v$ term.} Notice that in our case the gauge potential is fixed, since the Carroll structure of the horizon is actually embedded in the Kerr--Newman spacetime. Therefore we are not free to re-scale the gauge potential with powers of $\rho$, as to obtain an \emph{electric} contraction: such a rescalling would violate the Einstein equations. For this reason, the only consistent Carroll-covariant electromagnetism is given by the \emph{magnetic} contraction.

The magnetic field can be consistently calculated from Eq.~\eqref{B2}, where \mbox{$F_{AB}=2\partial_{[A}A_{B]}$}, obtaining directly 
\begin{equation}\label{eq:BintheHorizon}
    B = \frac{2 a Qr_+\cos\theta}{(r_+^2+a^2\cos^2\theta)^2}\,,
\end{equation}
which is the same result as taking the projection of the magnetic field onto the horizon in Eq.~\eqref{eq:Our-B}. This should not be surprising because the $\rho\to0$ limit \emph{is} essentially a suitable way to project into the horizon such that the horizon structure is manifest. Actually the geodesic embedding of the horizon is just a way of writing the embedded surface in Gaussian null coordinates.

\begin{figure}[h!]
    \centering
    \includegraphics[
    width=8.6cm]{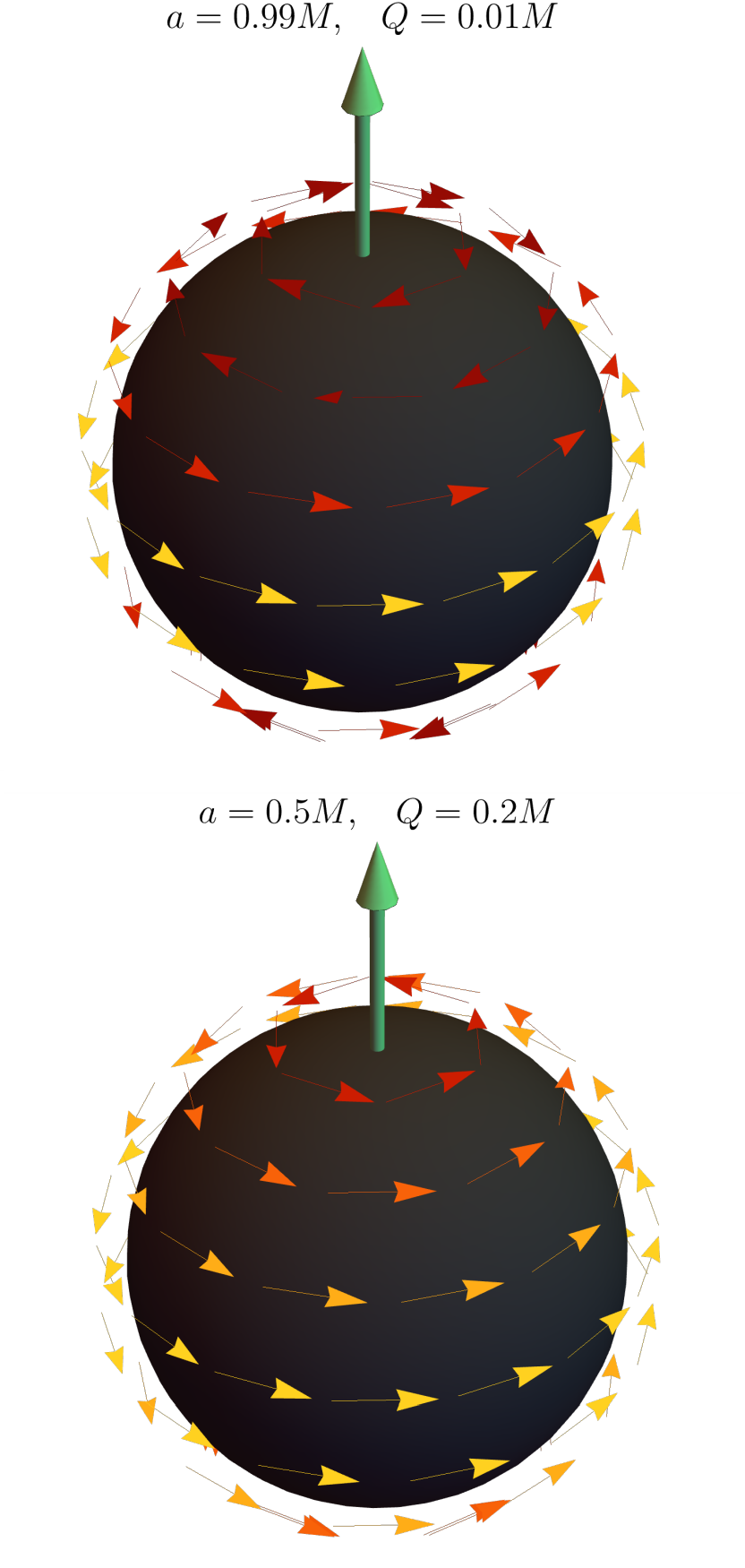}
    \caption{{\bf Carrolian motion at the horizon.} Schematic representation
    of the velocity field of the Carroll particle with $\mu\chi/\kappa_{\rm mag}=1$ for two different black holes, as indicated in the figure. The colour of the arrows indicates the magnitude of the velocity, with brighter colours corresponding to larger values. The green arrow indicates the direction of angular momentum of the black hole. }
    \label{fig:Rotations}
\end{figure}

\begin{figure}
    \centering
    \includegraphics[width = 0.9\columnwidth]{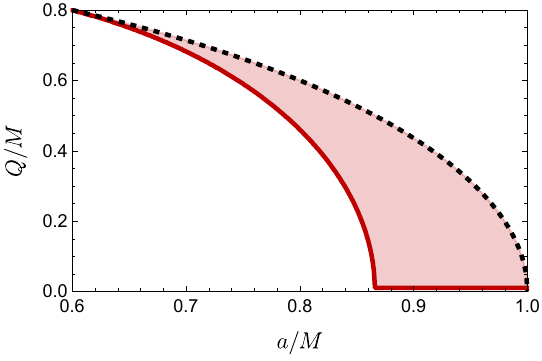}
    \caption{Region in the parameter space $(a/M, Q/M)$ where the exotic particles rotate in opposite directions in different regions of the horizon. The black dashed line represents an extremal black hole.}
    \label{fig:Change}
\end{figure}

Going back to the motion of a Carrollian particle in the horizon, its equations of motion can be read directly from Eq.~\eqref{eq:EoM}:
\begin{equation}
    \dv{\theta}{v} = 0, \quad \dv{\varphi}{v} = \frac{2Q\mu \chi}{\kappa_{\rm mag}} \frac{ar_+}{\Sigma_+^3 (r_+^2 + a^2)}\left(r_+^2 - 3a^2 \cos^2\theta\right).
\end{equation}
These exotic particles would exhibit a circular azimuthal motion at the horizon, see Fig~\ref{fig:Rotations}. When the spin and charge parameters are both small then the particle would co-rotate with the spin of the black hole depending on the relative sign of the magnetic moment $\mu$ and the charge of the black hole $Q$. However, if the black hole is sufficiently close to extremality, the direction of the rotation changes when moving closer to the poles (an interesting effect not discussed in \cite{main,main2}). The range of parameters where this effect happens is shown in Fig.~\ref{fig:Change}.
In particular, that implies that there are particular points at the horizon where the exotic particles would not move due to the vanishing gradient of the induced magnetic field. 

We show explicitly this behaviour in Fig.~\ref{fig:Rotations}. This allows for an interesting thought experiment: two exotic photons with opposite magnetic moments could be placed to co-rotate at very close latitudes within the horizon due to the same magnetic field. This could lead to intriguing effects if the particles were allowed to couple to each other, since it was shown that despite the ultrarrelativstic limit, Carrollian particles can retain non-trivial interactions~\cite{Bergshoeff:2014jla}.

\section{Assisted Anyonic spin-Hall effect}\label{sec:assisted}

There is another example of an (astrophysically well motivated) black hole horizon where we can have non-trivial Carrollian motion -- that of a (possibly weakly charged) black hole in a uniform magnetic field~\cite{wald:1974np}.

In a test field approximation, the corresponding vector potential can be obtained from the underlying isometries of the Kerr solution. Namely,
consider a spacetime with a Killing vector $\zeta$, i.e. 
\begin{equation}
    \nabla_\mu \zeta_\nu +\nabla_\nu \zeta_\mu=0\,.
\end{equation}
Then, expressing the second derivative of the Killing field in terms of the Riemann tensor and using the Bianchi identity we have the following integrability condition:
\begin{equation}\label{eq: Int Con}
    \nabla_\mu\nabla_\nu\zeta_\rho =\zeta^\sigma R_{\sigma\mu\rho\nu}\,,
\end{equation}
where $R_{\sigma\mu\nu\rho}$ is the Riemann tensor. Identifying the gauge field with the Killing vector,
\begin{equation}\label{jak}
A_\mu\propto\zeta_\mu \quad \Leftrightarrow \quad    F_{\mu\nu}\propto\nabla_\mu \zeta_\nu=\nabla_{[\mu} \zeta_{\nu]}\,, 
\end{equation}
we thus have, $dF=0$, and 
\begin{equation}
\nabla_\mu F^{\mu\nu}=R_{\mu}{}^\nu \zeta^\mu\,.
\end{equation}
That is, in vacuum (where $R_{\mu\nu}=0$), $\zeta$ gives rise to the solution of vacuum Maxwell equations via~\eqref{jak}.

Following Wald~\cite{wald:1974np} we  apply this to the Kerr spacetime \eqref{eq:KNmetric}, where $\Delta=r^2+a^2-2Mr$. Choosing $B_0$ to represent the strength of the constant magnetic field surrounding the black hole and $Q_0$ to represent a small charge for the black hole, one finds the following
vector potential:
\begin{equation}
    A =\frac{1}{2}B_0\left((1-2a\Omega_+)\partial_\varphi +2a\partial_v \right) -\frac{Q_0}{2M} (\partial_v-a\Omega_+\partial_\varphi)\,,
\end{equation}
upon employing the two Killing vectors of the Kerr geometry $\partial_v-\Omega_+ \partial_\varphi$ and $\partial_\varphi$. Lowering the index and performing a simple gauge transformation, we thus have:
\begin{align}
    A =&  - \frac{Q_0r}{\Sigma}\left(\frac{\Sigma_+}{r_+^2 + a^2}\dd v - a\sin^2\theta\,\dd \varphi\right)\,\nonumber\\
    &-\frac{B_0(1+\cos^2\theta)a\Delta}{2\Sigma} \left(\frac{\Sigma_+}{r_+^2 + a^2}\dd v -a \sin^2\theta\: \dd \varphi + \dd r\right)\nonumber\\
    & -\frac{B_0 (r^4-a^4)}{2 \Sigma}\dd \varphi\,.
\end{align}
Notice that the electromagnetic potential can then be written as Eq.~\eqref{eq:A} for a charged black hole added to a new term associated with the external magnetic field $B_0$.

One can repeat the procedure outlined in Section \ref{sec: Ind Car Mot} to find the induced magnetic field driving the exotic Carrollian motion. In fact one finds the identical expression but with a renormalized charge/field magnitude: 
\begin{equation}\label{Bind}
    B=\frac{\left(B_0 \left(r_+^4-a^4\right) +2a Q_0 r_+\right)\cos\theta}{\left(r_+^2+a^2 \cos ^2\theta\right)^2}\,.
\end{equation}
Notice that the test charge enters into this expression in the same way as the charge in the Kerr--Newman case. 

Equation~\eqref{Bind} allows us to interpret the anyonic spin-Hall effect in terms of the effective magnetic field seen by Carrollian particles in the black hole. Namely, let $B_Q$ denotes the  constant that characterizes the magnetic field induced by the black hole's charge $Q$:
\begin{equation}
    B_Q = \frac{2aQr_+}{r_+^4 - a^4}\,.
\end{equation}
Then, the total induced magnetic field in the horizon of a weakly charged rotating black hole in an external magnetic field \eqref{Bind} can be written as
\begin{equation}
    B=\frac{ (B_Q+ B_0)\left(r_+^4-a^4\right)\cos\theta}{\left(r_+^2+a^2 \cos ^2\theta\right)^2}\,.
\end{equation}

That is, the dynamics of the Carroll particle will be the same as the one found for a Kerr--Newman black hole, but with the above change in the corresponding magnetic field, $B_Q \rightarrow B_Q +B_0$. It is then possible to both increase the anyonic spin-Hall effect, or to decrease it, by using an external $B_0$ field. For this reason we call this the assisted anyonic spin-Hall effect. Notice that by choosing a magnetic field opposite to the direction of rotation of the black hole, it is also possible to completely cancel the effect, and to make the Carroll particles stand still in the horizon regardless of their charge or spin. 

Moreover, it is interesting that this induced magnetic field in the horizon will survive even in the limit of zero rotation of the black hole. That is, if one considers a {\em Schwarzschild black hole} immersed in an (asymptotically  constant) magnetic field, the exotic Carrollian motion will persist on its horizon. Taking the $a\to 0$ limit, the magnetic field becomes simply 
\begin{equation}
B= B_0\cos\theta\,.    
\end{equation}
An exotic photon like the one considered before would then exhibit a circular motion with period $T = 2\pi\kappa_\text{mag} r_s^2 / (B_0\mu\chi)$ where $r_s = 2M$ is the Schwarzschild radius. Notice that this period is independent of the position of the particle on the horizon. 

On the other hand, for the {\em extremal Kerr} black hole, we have $r_+=a$. It is well known that in that case the external magnetic field is expelled from the horizon -- the so called {\em black hole Meissner effect}, e.g. \cite{bivcak1976stationary, Chamblin:1998qm, Penna:2014aza, Bicak:2015lxa}. In such a case, the contribution of $B_0$ drops out in \eqref{Bind},  and only the charged induced magnetic field contributes, 
\begin{equation}
B=\frac{2Q\cos\theta}{a^2(1+\cos^2\!\theta)^2}\,.    
\end{equation} 

\section{Effect of a magnetic monopole}\label{app:magneticCharge}
Let us finally study the effect of a magnetic charge on the motion of Carrollian particles. In the presence of a magnetic monopole $Q_m$, the Kerr--Newman metric takes the same form of~\eqref{eq:KNmetricNew}, with the replacement of the function $\Delta$ by 
\begin{equation}
\Delta=r^2+a^2-2Mr+Q_e^2+Q_m^2\,,
\end{equation} 
while the vector potentials extends as
\begin{align}
    A=&-\frac{Q_er}{\Sigma}\left(\frac{r_+^2 +a^2\cos^2\theta}{r_+^2 +a^2}\dd v-a\sin^2\!\theta \dd\varphi\right)\nonumber\\&+\frac{Q_m \cos \theta}{\Sigma}\left(a\left(\frac{r^2-r_+^2}{r_+^2+a^2}\right)\dd v + (r^2 + a^2)\dd\varphi\right)\,,
\end{align}
where we now use $Q_e$ to denote the electric charge. In this case, we find the following induced magnetic field:
\begin{equation}
B= \frac{2Q_e ar_+\cos\theta - Q_m (r_+^2-a^2\cos^2\theta)}{(r_+^2 + a^2 \cos^2\theta)^2}\,.
\end{equation}

\begin{figure}[t]
    \centering
    \includegraphics[width = 0.9\columnwidth]{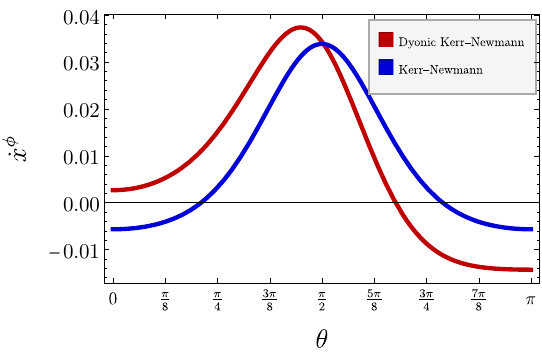}
    \caption{{\bf Effects of magnetic charge.} 
    Here we plot a comparison of the Carrollian velocity in the $\varphi$ direction generated by the induced magnetic fields at the horizon. The velocity for ordinary Kerr--Newman is shown in blue and in red for magnetically charged (dyonic) Kerr--Newman. We choose the same electric charge $Q_e=0.1M$ and $a=0.95M$ for both plots, while in the case of magnetic charge we pick $Q_m=Q_e/2$. These are such that the black holes are not extremal. One can see the presence of a magnetic monopole causes a phase shift in velocities and an asymmetry between the north and south poles. 
    }
    \label{fig:Change2}
\end{figure}

The appearance of the new term  qualitatively modifies the induced motion on the horizon. Namely, the presence of a magnetic monopole causes a phase shift in velocities and an asymmetry between the north and south poles, as depicted in Fig.~\ref{fig:Change2}.

\section{Conclusions}\label{sec:conclusions}

We have studied the motion of massless particles with anyonic spin and no charge in the horizon of a Kerr--Newman black hole. Through our analysis we found that the magnetic field previously prescribed in~\cite{main,main2} is not compatible with the Carrollian structure induced at the black hole horizon, and thus is inconsistent with the equations of motion in a Carroll manifold~\cite{first}. We then computed the Carroll invariant magnetic field induced in the horizon and obtained the corresponding equations of motion for a Carroll particle. The resulting magnetic field is smaller in magnitude than the one previously found, but points in the same direction, so that massless anyonic particles rotate around the axis of rotation of the black hole, displaying the anyonic spin-Hall effect described in~\cite{main,main2}. 
Interestingly, for `near extremal' black holes, there 
are regions close to the poles on the horizon where the Carrollian particles counter-rotate to the particles at lower latitudes, the two being separated by a critical latitude (a transition region) where the particle remains at rest. Using this effect, it would be tempting to study the interaction of two particles with opposite magnetic moments that co-rotate at close latitudes near the critical latitude.

We also studied the case where a weakly charged Kerr--Newman black hole is immersed in an external magnetic field. Exploring this case allowed us to define a constant, $B_Q = 2aQr_+/(r_+^4 - a^4)$, which characterizes the induced magnetic field in the horizon of a charged black hole with charge $Q$. We then showed that the effect of the external magnetic $B_0$ field is to change the magnetic field $B_Q$ to $B_Q+B_0$. Thus, if the magnetic field points opposite to the axis of rotation of the black hole, it might prevent the motion of the Carroll particle or generate motion against the direction of angular momentum. Overall, we have characterized the assisted anyonic spin-Hall effect in black hole horizons.

Finally, we have extended our considerations by allowing a presence of the magnetic charge in rotating black hole spacetimes.  This leads to a qualitatively different picture, where the Carollian velocity is no longer symmetric between the north and south poles and a phase shift in velocities is observed. 


For future studies, it would also be interesting to study the motion of Carrolian particles induced by other magnetic fields around black holes. For example, to consider the fully back-reacting Ernst solution~\cite{Ernst:1976mzr}, the fields of various sources outside of the black hole~\cite{bivcak1976stationary},  solutions of force-free electrodynamics~\cite{Brennan:2013kea}, 
or the magnetic fields of cosmic strings piercing the black hole horizon~\cite{Gregory:2013xca}.
One may also be tempted to look at the cases where (very strong) magnetic fields feature deviations from Maxwell's theory  and are described by non-linear electrodynamics, e.g.~\cite{Sorokin:2021tge}. Perhaps the most important open question is whether the demonstrated motion of Carrollian particles in the horizon, despite being an interesting mathematical possibility, could produce some observational features that could (at least in principle) be measured in future experiments. We leave this important question for future studies.

\acknowledgements

{The authors thank L.Marsot, P.-M. Zhang, and P. A. Horvathy for discussing their method of obtaining the resulting magnetic field in the horizon of a Kerr black hole.} F. G. and T. R. P. acknowledge support from the Natural Sciences and Engineering Research Council of Canada (NSERC) via the Vanier Canada Graduate Scholarship. J. R-Y. is supported by VILLUM FONDEN (grant no. 37766), by the Danish Research Foundation, and under the European Union’s H2020 ERC Advanced Grant “Black holes: gravitational engines of discovery” grant agreement no. Gravitas–101052587. Research at Perimeter Institute is supported in part by the Government of Canada through the Department of Innovation, Science and Industry Canada and by the Province of Ontario through the Ministry of Colleges and Universities. 
{Perimeter Institute and the University of Waterloo are
situated on the Haldimand Tract, land that was promised
to the Haudenosaunee of the Six Nations of the Grand
River, and is within the territory of the Neutral, Anishnawbe, and Haudenosaunee peoples.}

{\color{blue}

}

\bibliography{references.bib}

\end{document}